# Interaction between superconductor and ferromagnetic domains in iron sheath: peak effect in MgB$_2$/Fe wires

J. Horvat[a], W. K. Yeoh, and L. M. Miller

ISEM, University of Wollongong, NSW 2522, Australia

Interaction between the superconductor and ferromagnet in MgB$_2$/Fe wires results in either a plateau or a peak effect in the field dependence of transport critical current, $I_c(H)$. This is in addition to magnetic shielding of external field. Current theoretical models cannot account for the observed peak effect in $I_c(H)$. This paper shows that the theoretical explanation of the peak effect should be sought in terms of interaction between superconductor and magnetic domain structure, obtained after re-magnetization of the iron sheath by the self-field of the current. There is a minimum value of critical current, below which the re-magnetization of the iron sheath and peak effect in $I_c(H)$ are not observed.

Measurements of transport critical current ($I_c$) of MgB$_2$ superconducting wires sheathed with iron revealed an improvement of $I_c$ that was stronger than expected from a simple magnetic shielding by the sheath[1]. Instead of a monotonous decrease of $I_c$ with field obtained for the copper sheathed MgB$_2$ wires[2], the use of the iron sheath results in appearance of a plateau in the field dependence of $I_c$ at high temperatures and of a "peak effect" for temperatures lower than about 27K [3]. Mere magnetic shielding by the sheath cannot account for these results[1,3]. This effect is of interest for development of MgB$_2$ wires, because iron seems to be a material of choice as a sheath for MgB$_2$ cores[4-8].

Due to large value of $I_c$ for MgB$_2$ wires, the transport measurements of $I_c$ are performed by pulsed current method to avoid heating of the sample. This raises a possibility that the peak effect in $I_c(H)$ is an artifact of the pulsed current method. However, combining the pulsed current measurements at low fields and dc current measurements at high fields, as well as measuring the copper-sheathed MgB$_2$ wires, it was shown that the peak effect is indeed caused by the iron sheath[2]. This paper presents experimental results that point to a mechanism for improvement of $I_c(H)$ that has still not been considered in the theoretical models. It is hoped that these results will prompt development of the models capable of explaining the observed peak in $I_c(H)$ caused by the iron sheath and in turn help employ this effect for tailoring the properties of superconducting wires.

The interaction between a superconductor and ferromagnet has been a focus of research in recent years, resulting in several new models[9-15]. The explanation for the unusually strong improvement of $I_c$ by the iron sheath was initially sought in terms of the overcritical state model[12-16]. However, the overcritical state model is applicable only to the samples in the form of thin strips[12-14] placed in magnetic field, whereas our measurements are performed with the transport current through samples of cylindrical shape placed in external field H. Moreover, the observed peak effect in $I_c(H)$ [2,3] is not predicted in this model. The reliability of magnetic measurements that seem to support this model[16] is also questionable, because the effects of the cavities in the sample and microcracks introduced by removing the iron sheath were not taken into account[17,18] and no transport current was used in Ref.[16].

---

[a] Electronic mail: jhorvat@uow.edu.au

Kovač et al.[19] performed transport dc measurements of $I_c(H)$ for Bi2223/Ag tapes before and after an iron sheath was mounted around the tapes. Their results were consistent with our measurements of $MgB_2$/Fe wires[1], showing that the observed effect is not limited to $MgB_2$ superconductor. They also proposed a model for this effect, based on addition of the external field and self-field of the sample[20]. The model was in good agreement with the measured plateau in $I_c(H)$ at high temperatures. However, their model could not describe the peak effect in $I_c(H)$ at low temperatures, where $I_c$ is higher. Further testing of their model showed that the effects of the addition of the self-field to the external field are not responsible for the peak effect[2].

In this paper, we present results that pinpoint the mechanism for occurrence of the peak effect in $I_c(H)$. These results are obtained for a large number of samples measured over the last four years as $MgB_2$ wires were developed at ISEM. The samples were round iron-sheathed wires, with either pure $MgB_2$, SiC, or carbon-nanotube doped $MgB_2$. The value of transport $J_{c0}$ was of the order of $10^6$ A/cm$^2$ at 20K. The diameters of the superconducting core and iron sheath were typically about 0.8mm and 1.2mm, respectively. The samples were prepared by the powder-in-tube method. Detailed description of the sample preparation can be found elsewhere[4,8,21]. The resulting samples contained less than 5% of MgO, with $MgB_2$ being the only superconducting phase. Their critical temperatures were 38-39K, as obtained from the measurements of ac susceptibility. $I_c$ was obtained from voltage-current (V-I) characteristics, using the pulsed current source and four-probe method and H was perpendicular to the long axis of the wire. The field H was constant during the measurements of each V-I characteristic, whereas current through the sample was increased from zero to its peak value in $10^{-3}$ seconds. The fast-changing voltage on sample was recorded by a digital oscilloscope.

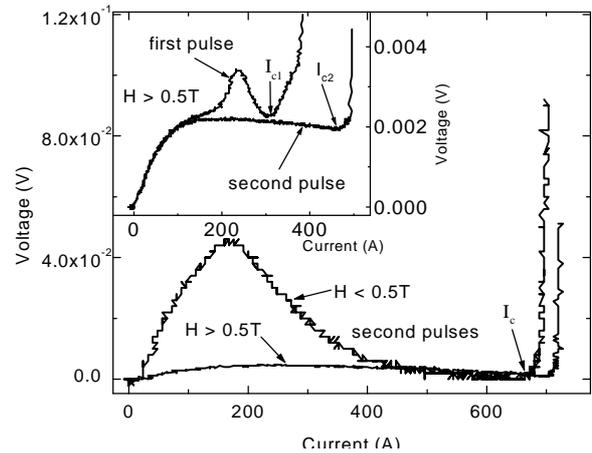

Figure 1: Typical voltage-current characteristics for an iron-sheathed $MgB_2$ wire at two different fields H, for second curent pulse after changing H. The peak at low currents occurs for H < 0.5T. Inset: typical V-I for H > 0.5T. The peak at low currents often occurs in the first pulse of the current after H was changed. There is no peak for the subsequent current pulses.

A typical V-I characteristic for small external field H is shown in Fig.1. Because the current through the sample varies in time, its self-field induces a background signal in the voltage taps of the sample. In addition, a strong change of magnetization of the iron sheath is induced by the self-field. This results in a peak in V-I for low values of the current in each current pulse. Once the iron sheath is almost fully magnetized by external field H, this peak disappears, typically above 0.5T. Even though the value of the background voltage induced by the self-field in the V-I characteristics is of the order of 1 mV, accurate values of $I_c$ can still be obtained for $MgB_2$ superconductor because of sharp increase of the voltage as the current reaches $I_c$. Thanks to this, the pulsed current measurements performed at low fields are in very good agreement with the dc current measurements performed at high fields[2]. However, accurate measurements of $I_c$ cannot be performed for high H with pulsed current method, because the V-I characteristics of $MgB_2$ are less sharp at high fields.

There is an unusual feature of V-I characteristics for T < 29K, for fields at which the peak effect occurs. The value of $I_c$ obtained with the first pulse of the current after changing the field H is lower than the one obtained in all subsequent pulses, with H kept constant (inset to Fig.1). Additionally, there are often one or more peaks in V-I for the first pulse (inset to Fig.1). As opposed to the peak in V-I for H < 0.5T (Fig.1), this peak does not appear any more in the second and subsequent pulses.

Figure 2 shows a typical field dependence of $I_c$ for round iron-sheathed $MgB_2$ wire at 20K. The open symbols show $I_c$ obtained in the first current pulse after H was changed, $I_{c1}$. There is a significant scattering of experimental points, with occasional jumps in the data. The values of $I_{c1}$ are not reproducible: another set of measurements generally gives different, highly scattered points in $I_{c1}(H)$. $I_{c1}$ is a result of transient effects in the sample, occurring in the initial $\sim 10^{-3}$ s of the first current pulse. The solid symbols in Fig.2 show the critical current obtained in the second current pulse ($I_{c2}$). This value of $I_c$ remains the same in the third or any subsequent pulse. Changing the direction of field H into the opposite (squares) does not change the value of $I_{c2}$. An important feature of Fig.2 is that the values of $I_{c1}$ and $I_{c2}$ overlap at high fields.

Figure 3 shows typical $I_c(H)$ for an iron sheathed $MgB_2$ wire at 11, 20 and 24K. $I_{c1}$ and $I_{c2}$ are shown by open and solid symbols, respectively. This figure shows that the difference between $I_{c1}$ and $I_{c2}$ and the peak effect are obtained only if $I_{c2}$ exceeds a particular value, $I_m$. The value of $I_m$ in Fig.3 was about 215, 225 and 240A for temperatures of 24, 20 and 11K, respectively. This difference in $I_m$ is just on the limit of the experimental uncertainty.

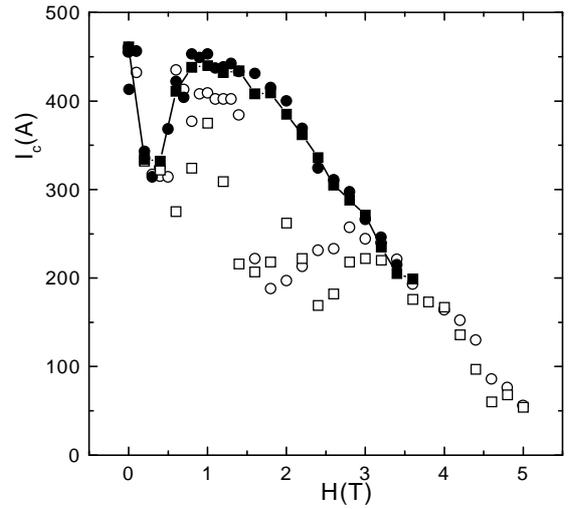

Figure 2: Field dependence of $I_c$ for carbon-nanotube doped $MgB_2$/Fe wire at T = 20K. Open symbols are obtained with the first current pulse after changing field H and solid symbols are obtained with second and all subsequent pulses. The round and square symbols are for positive and negative H, respectivelly.

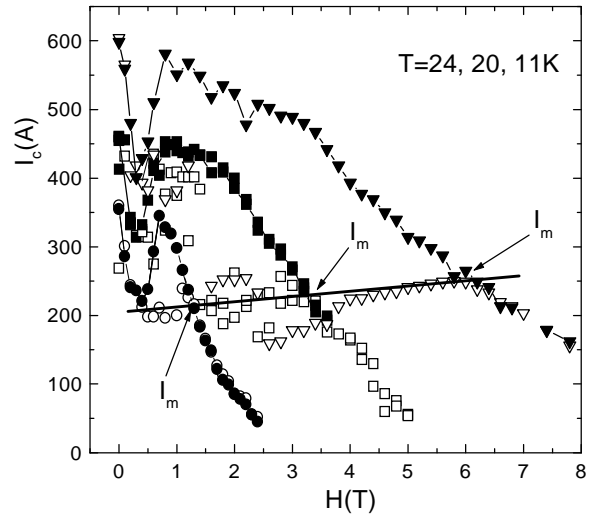

Figure 3: Field dependence of $I_c$ for a carbon-nanotube doped $MgB_2$/Fe wire at T = 24, 20 and 11K, shown with round, square and triangle symbols, respectivelly. Open symbols are obtained with the first current pulse after changing field H ($I_{c1}$) and solid symbols are obtained with second and all subsequent pulses ($I_{c2}$). $I_m$ is the current below which $I_{c1}$ and $I_{c2}$ start overlapping.

The values of $I_{c1}$ tend to aggregate around the line that connects $I_m$ for different temperatures in Fig.3. The difference between $I_{c1}$ and $I_{c2}$ decreases with increasing temperature and it finally disappears for temperatures at which $I_c < I_m$ for all fields. This temperature is around 29K for our samples. For T > 29K, $I_c(H)$ exhibits a plateau instead of a peak[1,3]. The same qualitative results were obtained for all the samples measured over the last four years, even though the value of $I_{c0}$ of these samples varied. Kovač et al. also suggested that a peak in $I_c(H)$ for Bi2223/Ag tapes sheathed with iron occurs only for large enough values of $I_c$[19]. This would imply that the presented results probably apply for a variety of different superconductors.

To verify the necessity that $I_c > I_m$ to obtain the difference in $I_{c1}$ and $I_{c2}$, we measured V-I with a series of pulses, starting from a low value of the peak current in the pulse and gradually increasing this value in subsequent pulses. For example, measuring V-I at T = 20K, H = 0.9 T and the peak current in the first pulse of 220A, the current did not reach $I_c$. In the second pulse, the peak current was 280A and $I_c$ of 265A was obtained. In the third pulse, the peak current was again 280A, but $I_c$ was not reached. In the fourth pulse, the peak current was 520A and $I_c$ of 440A was obtained. In the fifth pulse with peak current of 520A, $I_c$ of 440A was obtained again. In another experiment, the first pulse had a peak current of 240A and $I_c$ was not reached. However, the second pulse with the peak current of 520A produced the $I_c$ of 410A, which corresponded to $I_{c2}$ for that sample. This and many other experiments showed that the peak current in the first pulse had to exceed about 230A (i.e. $I_m \approx 230A$) in order to obtain larger values of $I_c$ (i.e. $I_{c2}$).

We interpret our experimental results in terms of re-magnetization of the iron sheath by the self-field and interaction of the superconductor core with thus re-magnetized sheath. When the field H is changed, it magnetizes the iron sheath in perpendicular direction. Applying the current pulse with peak current higher than $I_m$, the iron sheath is re-magnetized by the self-field in the circular direction. This self-field has to exceed the coercive field of the iron to be able to induce the change of magnetization, which explains the experimentally observed requirement for $I_m$. The re-magnetization occurs in the first ~$10^{-3}$ seconds after the dc current is switched on, explaining why this effect is not observed in standard dc measurements using voltmeters.

The interaction of the superconductor with the iron sheath results in larger $I_{c2}$ only when the magnetic domain structure is magnetized by the circular self-field. This would result in the magnetic domains not having perpendicular component of magnetization at the iron/superconductor interface, except around the Bloch domain walls. When the field H is changed, the domains are magnetized in the perpendicular direction and there is substantial perpendicular component of magnetization at the iron/superconductor interface, resulting in a low critical current, $I_{c1}$. This state seems to be unstable, presumably because of limited local re-magnetization by the self-field taking place in the sheath. The observed plateau in $I_c(H)$ for T > 29K is reminiscent of $I_{c1}(H)$ and it probably occurs because $I_c < I_m$ at these temperatures[22].

There is currently no theory that would describe the interaction of type-II superconductor in the mixed state with the iron sheath of different domain configurations. The presented experimental results imply that such a theory would be able to account for the observed peak in $I_c(H)$ (Figs. 2 and 3), as well as for the plateau at higher temperatures[1,3].

**Acknowledgment**
This work was supported by Australian Research Council, under project DP0211240. W. K. Yeoh received an Australia-Asia Award from the Australian Government.